\documentclass{article}
\usepackage{spconf,amsmath,graphicx}
\usepackage{multirow}
\usepackage[caption=false,font=footnotesize]{subfig}

\title{JOINT BAYESIAN Gaussian Discriminant Analysis FOR SPEAKER VERIFICATION}
\name{Yiyan Wang, Haotian Xu, Zhijian Ou\thanks{This work is supported by NSFC grant 61473168.}}

\address{Speech Processing and Machine Intelligence (SPMI) Lab, Tsinghua University, Beijing, China \\
wangyiya14@mails.tsinghua.edu.cn, xht13@mails.tsinghua.edu.cn, ozj@tsinghua.edu.cn
}
\begin{document}
\ninept
\maketitle
\begin{abstract}

State-of-the-art i-vector based speaker verification relies on variants of Probabilistic Linear Discriminant Analysis (PLDA) for discriminant analysis.
We are mainly motivated by the recent work of the joint Bayesian (JB) method, which is originally proposed for discriminant analysis in face verification.
We apply JB to speaker verification and make three contributions beyond the original JB.
1) In contrast to the EM iterations with approximated statistics in the original JB, the EM iterations with exact statistics are employed and give better performance.
2) We propose to do simultaneous diagonalization (SD) of the within-class and between-class covariance matrices to achieve efficient testing, which has broader application scope than the SVD-based efficient testing method in the original JB.
3) We scrutinize similarities and differences between various Gaussian PLDAs and JB, complementing the previous analysis of comparing JB only with Prince-Elder PLDA.
Extensive experiments are conducted on NIST SRE10 core condition 5, empirically validating the superiority of JB with faster convergence rate and $9-13\%$ EER reduction compared with state-of-the-art PLDA.
\end{abstract}

\section{Introduction}
\label{sec:intro}

The state-of-the-art in speaker recognition is currently dominated by the i-vector approach, that models both speaker and channel variabilities in a single low-dimensional space termed the total variability subspace \cite{dehak2011front}.
An i-vector based speaker verification system mainly consists of three components, which are i-vector extractor based on Gaussian Mixture Models (GMMs) or Deep Neural Networks (DNNs) \cite{lei2014novel}, i-vector post-processing (e.g. length normalization) and discriminant analysis.
Note that this approach defers the decomposition of speaker and intersession variabilities to the stage of discriminant analysis, which is particularly important for this approach.



An attractive discriminant analysis technique is to construct likelihood ratio score based on probabilistic generative models such as the widely used Probabilistic Linear Discriminant Analysis (PLDA) \cite{prince2007probabilistic} with many variants.
Except the heavy-tailed PLDA \cite{kenny2010bayesian}, most variants are Gaussian, such as Prince-Elder PLDA~\cite{prince2007probabilistic}, Simplified PLDA (SPLDA)~\cite{garcia2014supervised}, Kaldi PLDA~\cite{kaldicode}, two-covariance model~\cite{Bruemmer2010}, and Ioffe PLDA~\cite{Ioffe2006a}.
Basically, Gaussian PLDA assumes that the $j$-th i-vector from speaker $i$ obeys the following decomposition \footnote{Throughout the paper, we assume the data have zero-mean after a standard centering preprocessing step and omit the global mean in the model.} :
\begin{displaymath}\label{equ:plda}
  x_{ij} = Fz_i + \epsilon_{ij}
\end{displaymath}
where the latent speaker factor $z_i$ and the intersession residual $\epsilon_{ij}$ are both Guassians and independently distributed. $F$ is the loading matrix spanning the speaker subspace.
Denote by $H_I$ the intra-personal hypothesis that one set of i-vectors $x_1$ and another set of i-vectors $x_2$ belong to the same speaker, and $H_E$ the extra-personal hypothesis that they are from different speakers.
The speaker verification problem can then be solved by thresholding the likelihood ratio score ${p(x_1,x_2|H_I)}/{p(x_1,x_2|H_E)}$.

The performance of PLDA largely depends on how it can be trained effectively to learn the within-class variability, which characterizes intersession residuals, and the between-class variability, which characterizes differences among speakers.
Some improvements include using data domain adaptation of PLDA parameters \cite{garcia2014supervised} and discriminative training \cite{Bousquet2016}.
In this paper, we are primarily concerned with addressing the two basic challenging issues for the current Gaussian PLDA family.
First, for PLDAs with subspace modeling, it is difficult to determine the subspace dimension which is crucial for performance. Low subspace dimension often leads to under-fitting, while high subspace dimension results in over-fitting.
Second, whether using subspace modeling or not, current PLDAs suffer from slow convergence of their implemented EM iterations.
As analyzed in\cite{chen2016efficient}, different parameterizations and selections of hidden variables in designing EM updates have significant effect on the convergence performance.
These basic issues hinder improving the performance of current PLDAs.


We are mainly motivated by recent the work of Joint Bayesian (JB) method \cite{chen2016efficient}, which was originally proposed for face verification.
In JB, there is no need to determine the subspace dimension, and it achieved faster convergence and more accuracy in \cite{chen2016efficient} for face verification.
We apply JB to speaker verification and make three main contributions.
1) We find that the EM updates with approximated 
statistics suggested in \cite{chen2016efficient} does not work in speaker verification problem. Instead, the EM iterations with exact statistics are employed and give better performance.
2) Inspired by Fisher LDA, we propose to do simultaneous diagonalization (SD) of the within-class and between-class covariance matrices to achieve efficient testing.
Compared to the SVD-based efficient testing method in \cite{chen2016efficient}, the new SD method can still be applied to reduce the testing complexity even in the case that the number of training samples per subject are different.
3) We scrutinize similarities and differences between various Gaussian PLDAs and JB, complementing the analysis of comparing JB only with Prince-Elder PLDA in \cite{chen2016efficient}. Moreover,
extensive experiments are conducted on NIST SRE10 core condition 5, empirically validating the superiority of JB in term of EM convergence rate and EER performance.


\section{Joint Bayesian Gaussian Discriminant Analysis}
\label{sec:jbl}

Joint Bayesian (JB) Gaussian discriminant analysis was first proposed in~\cite{chen2016efficient} for face verification. Its model formulation is similar to the two-covariance model~\cite{Cumani2013} but with different parameterizations and selections of hidden variables in EM training. For speaker verification, we use two independent Gaussians to represent speaker identity and intersession residuals respectively. The $j$-th i-vector of speaker $i$, denoted by $x_{ij} \in R^d$, is decomposed as:
\begin{displaymath}
\begin{split}
x_{ij}=\mu_i+\varepsilon_{ij}
\end{split}
\label{eq:4}
\end{displaymath}
where $\mu_{i}\sim\mathcal{N}(0,S_{\mu})$ is the speaker identity variable, $\varepsilon_{ij}\sim\mathcal{N}(0,S_{\varepsilon})$ models the within-speaker variability.
The model parameters are $\Theta=\{S_\mu,S_\varepsilon\}$.
The extracted $m_i$ i-vectors for speaker $i$ are denoted by $x_i=[x_{i1};\ldots;x_{im_i}]$.
The total hidden variables are stacked as  $h_i=[\mu_i;\varepsilon_{i1};\ldots;\varepsilon_{im_i}]$, which are Gaussian distributed with block diagonal covariance matrix $\Sigma_{h_i}=diag(S_\mu,S_\varepsilon,\ldots,S_\varepsilon)$. \par
The data likelihood for observed $x_i$ is
\begin{equation}\label{equ:jbfjoint}
  p(x_i)=\mathcal{N}(0,\Sigma_{x_i}),\Sigma_{x_i}=
  \begin{bmatrix}
    S_{\mu}+S_{\varepsilon} & S_{\mu} & \ldots & S_{\mu}\\
    S_{\mu} & S_{\mu}+S_{\varepsilon} & \ldots & S_{\mu}\\
    \vdots   & \vdots  & \ddots  & \vdots \\
    S_{\mu}       &S_{\mu}      & S_{\mu}      & S_{\mu}+S_{\varepsilon}
  \end{bmatrix}
\end{equation}
where the dimension of $\Sigma_{x_i}$ is $m_i\times d$.

The parameters $\Theta$ are estimated by the EM algorithm through iteratively optimizing the expected complete log-likelihood function as follows:
\begin{equation}\label{equ:expectedjb}
\begin{split}
\max_\Theta \sum_iE_{p(h_i|x_i;\Theta^{t})}[logp(h_i;\Theta^{t+1})]
\end{split}
\end{equation}
where $\Theta^{t}=\{S_\mu^t,S_\varepsilon^t\}$ are the parameters from the $t$-th EM update, and $\Theta^{t+1}$ the parameters to be updated in the $(t+1)$-th iteration.
Under this auxiliary objective function, the terms related to $S_\mu$ and $S_\epsilon$ are effectively decoupled, resulting in very elegant update equations for $S_\mu$ and $S_\epsilon$ \cite{chen2016efficient}.

In speaker verification testing, we calculate the log-likelihood ratio score to determine whether one set of i-vectors $x_1$ (including $m_1$ i-vectors) and another set of i-vectors $x_2$ (including $m_2$ i-vectors) are from the same speaker \footnote{By abuse of notation, here $x_1$ is not the data corresponding to speaker 1.} :
\begin{equation}
\begin{split}
r(x_1,x_2)&=logp(x_1,x_2)-logp(x_1) - logp(x_2)
\end{split}
\label{eq:7}
\end{equation}
Note that all the data likelihoods $p(x_1,x_2)$, $p(x_1)$ and $p(x_2)$ can be calculated through Eq.~\ref{equ:jbfjoint}, which involves matrix inversions.

To accelerate testing, \cite{chen2016efficient} employed SVD to obtain low rank approximations of the matrices appearing in the three log-likelihood terms in Eq.~\ref{eq:7}, which depend on $m_1+m_2$, $m_1$ and $m_2$ respectively.
Therefore, this speedup is more useful under the condition that the number of i-vectors is the same across all subjects, i.e. $m_i=m$. This is often satisfied in the task of face verification and face search.

Here we propose first to do simultaneous diagonalization (SD) of $S_\mu$ and $S_\varepsilon$, $\Phi^T S_\mu\Phi=K$ and $\Phi^T S_\epsilon \Phi=I$. Similar to Fisher LDA, we keep the first $s<d$ largest eigenvalues of $S_\mu^{-1} S_\varepsilon$, giving the low-rank diagonal matrix $K$. Denote by $\Phi$ the corresponding low-rank eigenvector matrix.
By defining $\Psi={\Phi^{-T}}$, we have $S_\mu=\Psi K \Psi^T$,  $S_\epsilon=\Psi\Psi^T$, and moreover,
\begin{displaymath}\label{equ:sd-covariance}
\Sigma_{x_i}=\Omega
\begin{bmatrix}
K+I & K & \ldots & K\\
K & K+I & \ldots & K\\
\vdots   & \vdots  & \ddots  & \vdots \\
K       &K      & K      & K+I
\end{bmatrix}
\Omega^T
\end{displaymath}
where $\Omega=diag(\Psi;\ldots;\Psi)$.
Based on this decomposition of $\Sigma_{x_i}$, the calculation of data likelihood $p(x_i)$ could be accelerated, if we take $\Omega$ to transform the i-vectors via pre-computation. The likelihood calculation then only involves inversion of diagonal matrices, reducing the complexity from $\mathcal{O}(d^3)$ to $\mathcal{O}(d)$.
Moreover, it can be seen that this speedup does not depend on $m_i$ and thus has broader applicability.
In this paper, we also conduct experiments to compare these two speedup methods over speaker verification accuracy.

\begin{table*}
	\vspace{-0.5cm}
	\centering
	\begin{tabular}{c| c| c|c|c}
		\hline
		Method  &   JB     &two-covariance  & SPLDA &   Kaldi PLDA    \\
		\hline
		Observation & \multicolumn{3}{c|}{$x_i=\{x_{ij},j=1,\ldots,m_i\}$} & $\bar{x}_i=\frac{1}{m_i}\sum_{j=1}^{m_i}x_{ij}$   \\
		\hline
		Model & \multicolumn{2}{c|}{$x_{ij}=\mu_i+\varepsilon_{ij}$} & $x_{ij}=Fz_i+\varepsilon_{ij}$ & $\bar{x}_i=\mu_i+\varepsilon_{i1}$  \\
		\hline
		$h_i$   & $\{\mu_i,\{\varepsilon_{ij}\}\}$ & $\{\mu_i\}$& $\{z_i\}$ & $\{\mu_i,\varepsilon_{i1}\}$ \\
		\hline
		EM objective function $Q(\Theta_{t},\Theta_{t+1})$   & $E_{p(h_i|x_i)}[logp(h_i)]$ & \multicolumn{2}{c|}{$E_{p(h_i|x_i)}[logp(x_i,h_i)]$}    & $E_{p(h_i|\bar{x}_i)}[logp(h_i)]$  \\
		\hline
		Subspace dimensionality setting & \multicolumn{2}{c|}{loose} & strict & loose\\
		\hline
		EM convergence   &  fast &  \multicolumn{2}{c|}{slow}    &fast \\
		\hline
	\end{tabular}
	\caption{The summary of the similarities and difference between JB, SPLDA, Kaldi PLDA and the two-covariance model.
		$x_{ij}$ denotes the $j$-th i-vector of speaker $i$.
		$\mu_i\sim\mathcal{N}(0,S_\mu)$ is the identity variable for speaker $i$, modeled by the between-class covariance $S_\mu$.  $\varepsilon_{ij}\sim\mathcal{N}(0,S_\epsilon)$ is the intersession residual, modeled by the within-class covariance $S_\epsilon$. For SPLDA, $z_i\sim\mathcal{N}(0,I)$ stands for the identity variable.}
	\label{table:com}
	\vspace{-0.2cm}
\end{table*}

\section{Connection with PLDA}

In this section, we investigate the connections between joint Bayesian\\(JB)~\cite{chen2016efficient}, Simplified PLDA (SPLDA)~\cite{garcia2014supervised} and Kaldi-PLDA~\cite{povey2011kaldi}.
We mainly show that different parameterization and selection of hidden variables lead to different behavior of the EM algorithm, and JB is superior to PLDAs in terms of EM convergence.
For the advantages of JB in allowing the data to implicitly determine the subspace dimensionality for maximal discrimination and favoring low-rank esimates of
$S_\mu$ and $S_\varepsilon$, the reader could refer to \cite{chen2016efficient}.
Table \ref{table:com} summarizes the similarities and differences between JB, SPLDA, Kaldi PLDA and the two-covariance model.

\subsection{Simplified PLDA (SPLDA)}

Basically, SPLDA \cite{garcia2014supervised} assumes that $j$-th i-vector from speaker $i$ obeys the following decomposition :
\begin{equation}\label{equ:original-plda}
x_{ij} = Fz_i + \epsilon_{ij}
\end{equation}
where the latent speaker factor $z_i\sim\mathcal{N}(0,I)$ and the intersession residual $\epsilon_{ij}\sim\mathcal{N}(0,\Lambda)$ are both Guassians and independently distributed. $F$ is the loading matrix spanning the speaker subspace.
The speaker subspace could be full rank, which is also known as the two-covariance model~\cite{Cumani2013}.

The parameters $\Theta=\{F,\Lambda\}$ \cite{garcia2014supervised} are estimated by the EM algorithm through iteratively optimizing the complete log-likelihood $logp(x_i,z_i;\Theta_{t+1})$ averaged over $p(z_i|x_i;\Theta_{t})$ where $\Theta_{t}=\{\Lambda_t,F_t\}$
\begin{equation}\label{equ:expectedplda}
\max_\Theta \sum_iE_{p(z_i|x_i;\Theta_{t})}[logp(x_i,z_i;\Theta_{t+1})]
\end{equation}

Different from Eq. \ref{equ:expectedjb} in JB, the hidden variables in SPLDA are only $z_i$'s,  excluding $\epsilon_{ij}$'s.\footnote{Including all hidden variables to derive the EM update for SPLDA is ill-posed under SPLDA's parameterization.}
Now we analyze the convergence property of the EM updates for SPLDA, analogous to \cite{chen2016efficient}. Note that maximizing Eq. \ref{equ:expectedplda} over $F_{t+1}$ is equivalent to minimizing over $F_{t+1}$
\begin{displaymath}
\begin{split}
\sum_i\sum_j trace({\Lambda}_{t+1}^{-1}E[(x_{ij}-{F}_{t+1}z_i)(x_{ij}-{F}_{t+1}z_i)^T])
\label{equ:16}
\end{split}
\end{displaymath}
It can be seen that
\begin{displaymath}\label{equ:splda-f-lambda}
\begin{split}
&E[(x_{ij}-{F}_{t+1}z_i)(x_{ij}-{F}_{t+1}z_i)^T] = \\
&(x_{ij}-F_{t+1}E[z_i])(x_{ij}-F_{t+1}E[z_i])^T\\
&+F_{t+1}(I-F_t^T(F_tF_t^T+\Lambda_{t})^{-1}F_t)F_{t+1}^T
\end{split}
\end{displaymath}
where
\begin{displaymath}\label{equ:splda-posterior}
\begin{aligned}
    E[z_i] &=F_{t}^T(F_{t}F_{t}^T+\Lambda_t)^{-1}x_{ij}\\
    E[z_iz_i^T]-E[z_i]E[z_i]^T &= I-F_t^T(F_tF_t^T+\Lambda_{t})^{-1}F_t
\end{aligned}
\end{displaymath}
When $\Lambda_t$ is small, by setting $F_{t+1}$ as $F_{t}$, we find that
\begin{displaymath}\label{equ:splda-first-term}
\begin{split}
x_{ij}-F_{t+1}E[z_i] \approx x_{ij}-F_{t+1}F_{t}^T(F_{t}F_{t}^T)^{-1}x_{ij}=0
\end{split}
\end{displaymath}
\begin{displaymath}\label{equ:splda-second-term}
\begin{split}
F_{t+1}(I-F_t^T(F_tF_t^T+\Lambda_{t})^{-1}F_t)F_{t+1}^T \approx \\ F_{t+1}(I-F_t^T(F_tF_t^T)^{-1}F_t)F_{t+1}^T=0
\end{split}
\end{displaymath}
Hence updating $F_{t+1}$ as $F_{t}$ approximately optimize the M-step and the EM-algorithm stalls upon a single iteration.

Note that theoretically the EM algorithm is only actually guaranteed to produce non-decreasing optimization of data likelihood through a series of parameter updates.
Strict convergence to local minima (or stationary points) requires further strong assumptions.
Combining this understanding of the EM algorithm and the above analysis of halt upon a single iteration, we could realize that the EM update for SPLDA could be easily stuck into a non-local minimum with small $\Lambda_t$.
The EM update for JB does not have such problem, since JB has different parameterization and selection of hidden variables. The faster convergence of the EM iterations for JB is also empirically observed in our experiments.

\subsection{Kalid PLDA}

The Kaldi is a widely used open-source speech recognition toolkit~\cite{povey2011kaldi}.
Here we examine the PLDA implementation in Kaldi code repository \cite{kaldicode}.
The conceptual starting point for Kaldi PLDA is the SPLDA model as shown in Eq.~\ref{equ:original-plda} with full rank $F$.
Next, Kaldi PLDA is only concerned with modeling the average i-vector for each speaker
$
\bar{x}_i = \sum_{j=1}^{m_i}x_{ij}/m_i
$,
which is distributed according to	
\begin{equation}\label{equ:mean-distribution}
  p(\bar{x}_i)=\mathcal{N}(0,FF^T+\frac{1}{m_i}\Lambda)
\end{equation}
where $m_i$ is the numbers of extracted i-vectors for speaker $i$.

Eq. \ref{equ:mean-distribution} is then treated as the data likelihood function. All the extracted i-vectors for each speaker are collapsed as a single sample - the average i-vector, which is assumed to obey the decomposition :
\begin{displaymath}\label{equ:decmpose-mean}
  \bar{x}_i = \mu_i + \bar{\varepsilon}_{i}
\end{displaymath}
where $\mu_i\sim\mathcal{N}(0,\Gamma)$ models the between-class variability with the covariance  $\Gamma=FF^T$ and the average residual 
$
\bar{\varepsilon}_{i} = \sum_{j=1}^{m_i}{\varepsilon}_{ij}/m_i \sim \mathcal{N}(0,\frac{1}{m_i}\Lambda)$ models the within-class variability.

The expected complete log-likelihood function for the EM algorithm is optimized to iteratively estimate $\Gamma$ and $\Lambda$, as follows:
\begin{displaymath}\label{equ:kaldi-q}
\max_\Theta\sum_iE_{p(\mu_i,\varepsilon_{i1}|x_i;\Theta_{t})}[logp(\mu_i,\varepsilon_{i1};\Theta_{t+1})]
\end{displaymath}
The parameterization of Kaldi PLDA is similar to JB, i.e. using two covariance matrices. Hence the EM iterations in Kaldi PLDA can also select the total hidden variables $(\mu_i,\varepsilon_{i1})$, with good convergence.
However, the additive decomposition only applies to the average i-vector in Kaldi PLDA. This is helpful for estimating between-class covariance but is detrimental for estimating within-class covariance.

At the testing phase, Kaldi PLDA also performs simultaneous diagonalization of $\Lambda$ and $\Gamma$. However, the significance of computational saving is less than the SD applied in JB, because JB calculates the joint likelihood of a number of i-vectors while Kaldi only calculates the likelihood of a single average i-vector.

\section{EXPERIMENTS}
\subsection{Dataset}
\label{ssec:subhead}
We conduct speaker verification experiments with different discriminant analysis techniques on the NIST SRE10 core condition 5, which includes 11982 speakers, 7169 target and 408950 nontarget trials \cite{martin2010nist}.  The DNN used in the experiments is trained on part of the Fisher data including about 600 hours of speech cuts. The i-vector extractor training data comprises 57517 speech cuts of 5767 speakers, which are from Switchboard, Fisher and NIST SRE 04, 06, and 08. Both JB and SPLDA are trained on SRE data, consisting of 36612 speech cuts and 3805 speakers from NIST SRE 04, 06, and 08.

\begin{table}[htb]
  \centering
  \scriptsize
  \begin{tabular}{l|c|c|c|c}
  \hline
                    & Fisher     & Switchboard    & SRE   & duration (hours) \\ \hline
  DNN-UBM           & $\surd$    &                &       & 600  \\
  i-vector extractor &  $\surd$  &   $\surd$      & $\surd$    &1890 \\
  SPLDA               &           &                 &$\surd$    & 1250   \\
  JB                &           &                 &$\surd$    &   1250      \\\hline
  \end{tabular}
  \caption{The data used to train the DNN-UBM, i-vector extractor, SPLDA and JB for speaker verification.}
  \label{table:3}
\end{table}
\vspace{-0.3cm}
\label{sec:typestyle}

\subsection{System Configuration}
\label{configuration}
The features used in the experiments are 40-dimensional Mel Frequency Cepstral Coefficients (MFCCs), including 20-dimensional static features and first-order derivatives. The speaker verification system uses a DNN-UBM with 5 hidden layers, 5335 senones, and a 600-dimensional i-vector extractor. The input of the DNN-UBM is the MFCCs extracted using 21 frames (11 frames before and 9 frames after). For discriminant analysis techniques, we implement three references namely LDA+COS, SPLDA and Kaldi PLDA. We apply the LDA to the i-vectors to obtain 200-dimension  features and use cosine distance metric for testing. For SPLDA, we set the dimension of the subspace to 300. For Kaldi PLDA, we use the default configuration of the Kaldi toolkit \cite{kaldicode}. The system performances are reported by equal error rate (EER) and minimum decision cost function (DCF) defined in NIST SRE08 and SRE10 \cite{martin2010nist}.

\subsection{Results}
\label{ssec:subhead}

\begin{figure}[htb]
	\small
	\vspace{-0.5cm}	
	\begin{minipage}[t]{0.5\linewidth}
		\centering
		\centerline{\includegraphics[width=5cm]{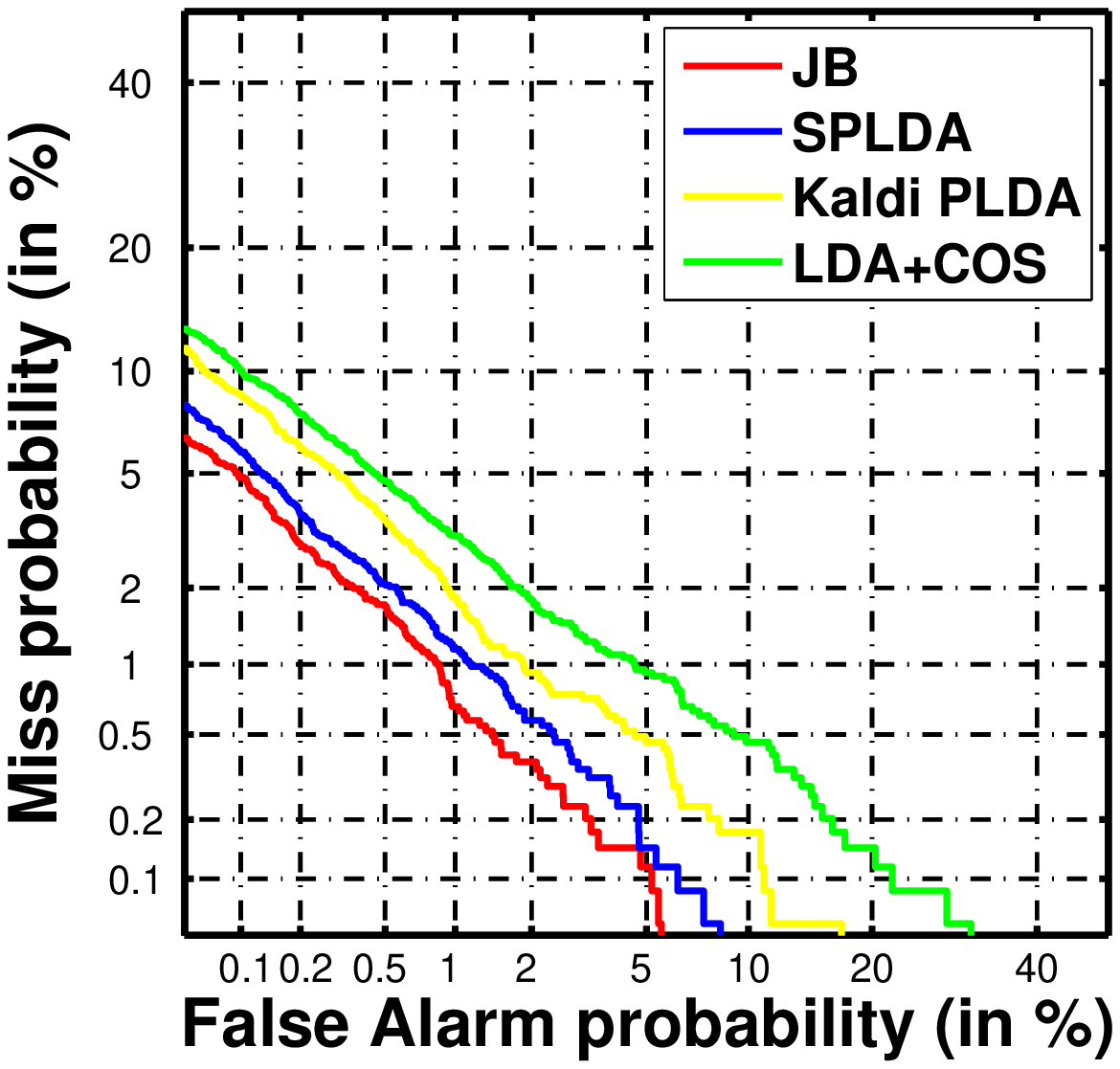}}
		\centerline{(a) SRE10 MALE}\medskip
		\vspace{-0.2cm}
	\end{minipage}
	\hfill
	\begin{minipage}[t]{0.5\linewidth}
		\centering
		\centerline{\includegraphics[width=5cm]{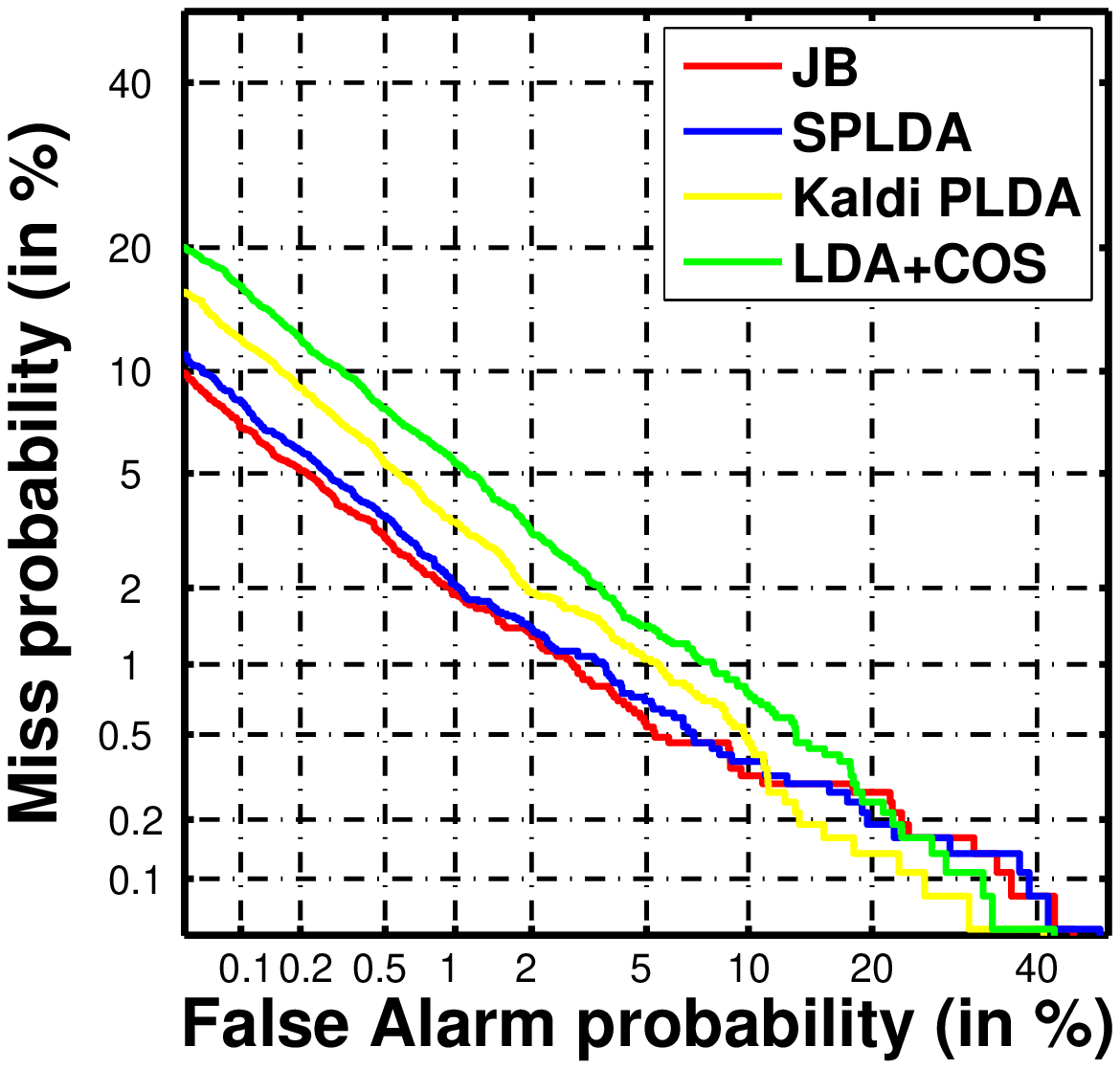}}
		\centerline{(b) SRE10 FEMALE}\medskip
		\vspace{-0.2cm}
	\end{minipage}
	
	\caption{DET curves for JB, SPLDA, Kaldi PLDA and LDA in SRE10 core condition 5 evaluation.}
	\label{fig:1}
	%
\end{figure}

\begin{figure}[htb]
	\begin{minipage}[b]{1.0\linewidth}
		\centering
		\centerline{\includegraphics[width=4.5cm]{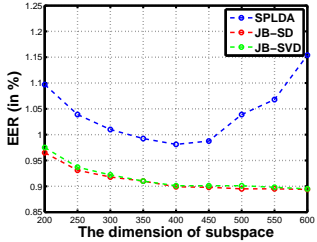}}
		\vspace{-0.2cm}
	\end{minipage}
	\caption{The influence of subspace dimensionality on JB and SPLDA using NIST SRE10 core condition male test data.}
	\label{fig:2}
	\vspace{-0.3cm}
\end{figure}

\subsubsection{Speaker Verification Performance}
\label{sssec:subsubhead}
We evaluate LDA+COS, Kaldi PLDA, SPLDA and JB on the NIST SRE10 core condition 5. These four models share the same training and test data. Fig. \ref{fig:1} illustrates the detection error trade-off (DET) curves of LDA+COS, SPLDA, Kaldi PLDA and JB for discriminant analysis with configurations described in Section~\ref{configuration}. From the results show in Tab. \ref{table:1}, we can conclude that :
\begin{itemize}
\item Compared to distance-based discriminant analysis LDA+COS, probabilistic generative model based methods such SPLDA, Kaldi PLDA and JB achieve better performance on EER.
\item In terms of EER, JB improves $13.0\%$ and $45.3\%$ compared to SPLDA and Kaldi PLDA respectively on SRE10 male tests, $9.2\%$ and $30.9\%$ on female tests. This verifies that JB with careful selection of hidden variables achieves better parameter estimation due to efficient EM updates.
\item SPLDA achieves better results on EER than Kaldi PLDA, because SPLDA utilizes the joint likelihood of i-vectors rather than the single average i-vector as used in Kaldi PLDA to estimate the parameters.
\end{itemize}

\subsubsection{Subspace Dimensionality}
\label{sssec:subsubhead}
Here we investigate the impact of the sub-space dimensionality of JB and SPLDA for discriminant analysis, which is shown in Fig. \ref{fig:2}. 
Two methods (SVD, SD) are used to reduce the subspace dimensionality of JB in speaker verification testing. It can be seen that: 1) The dimension of the subspace plays an important role in SPLDA  that lower may cause under-fitting, while higher may cause over-fitting. 2) The JB performance fluctuates slightly with the change of subspace dimension in testing, but the time complexity reduces from $O(d^3)$ to $O(s^3)$.  
Loose parameterization for JB makes it more robust since the dimension of the subspace is automatically fitted via data rather than manual defined. 3) SVD and SD have close performances but SD has wider applicability.

\subsubsection{Convergence Rate}
\label{sssec:subsubhead}
\begin{figure}
	\centering
	\vspace{-0.8cm}
	\hfil
	\subfloat[]{\includegraphics[width=0.51\linewidth]{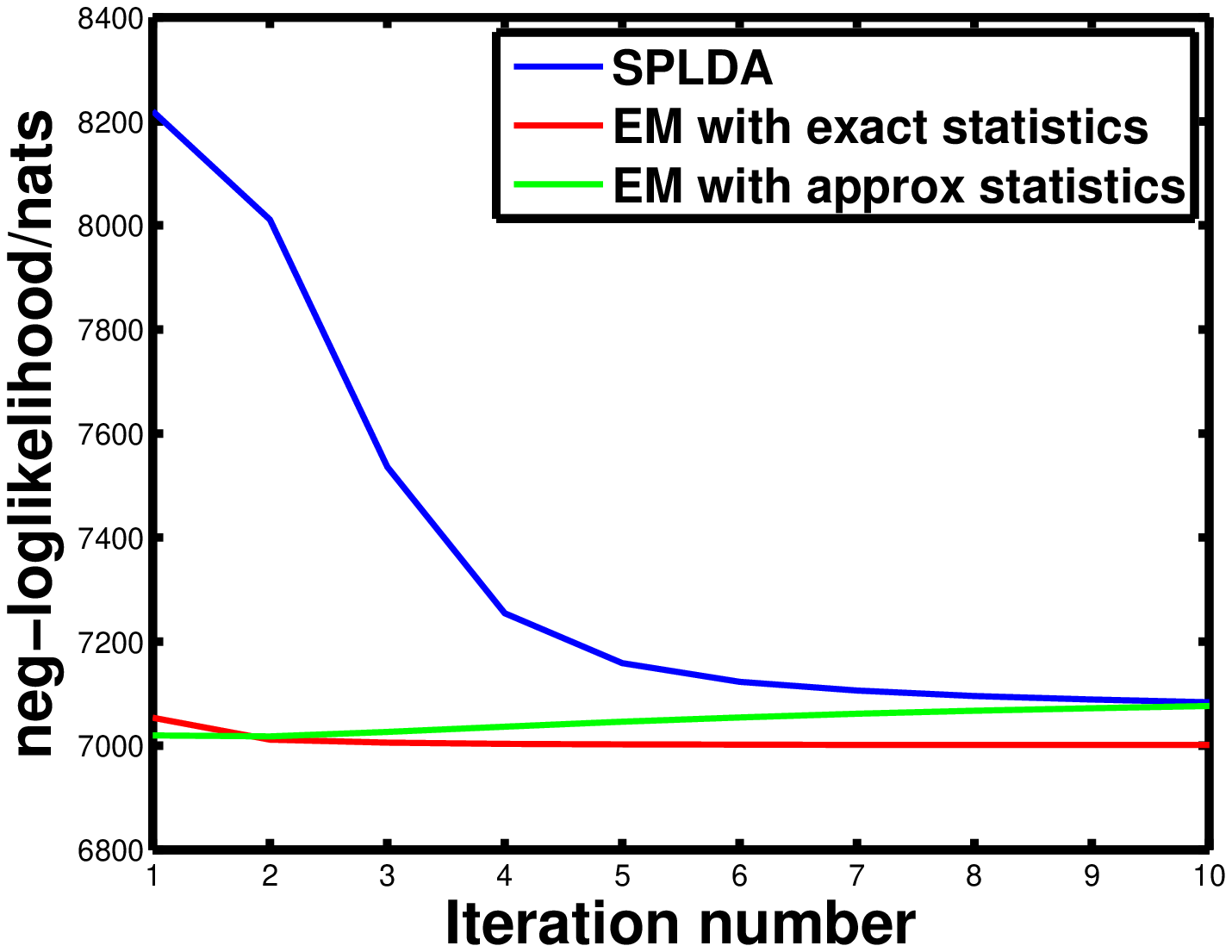}%
		\label{fig:exact-jbf-plda}}
	\subfloat[]{\includegraphics[width=0.51\linewidth]{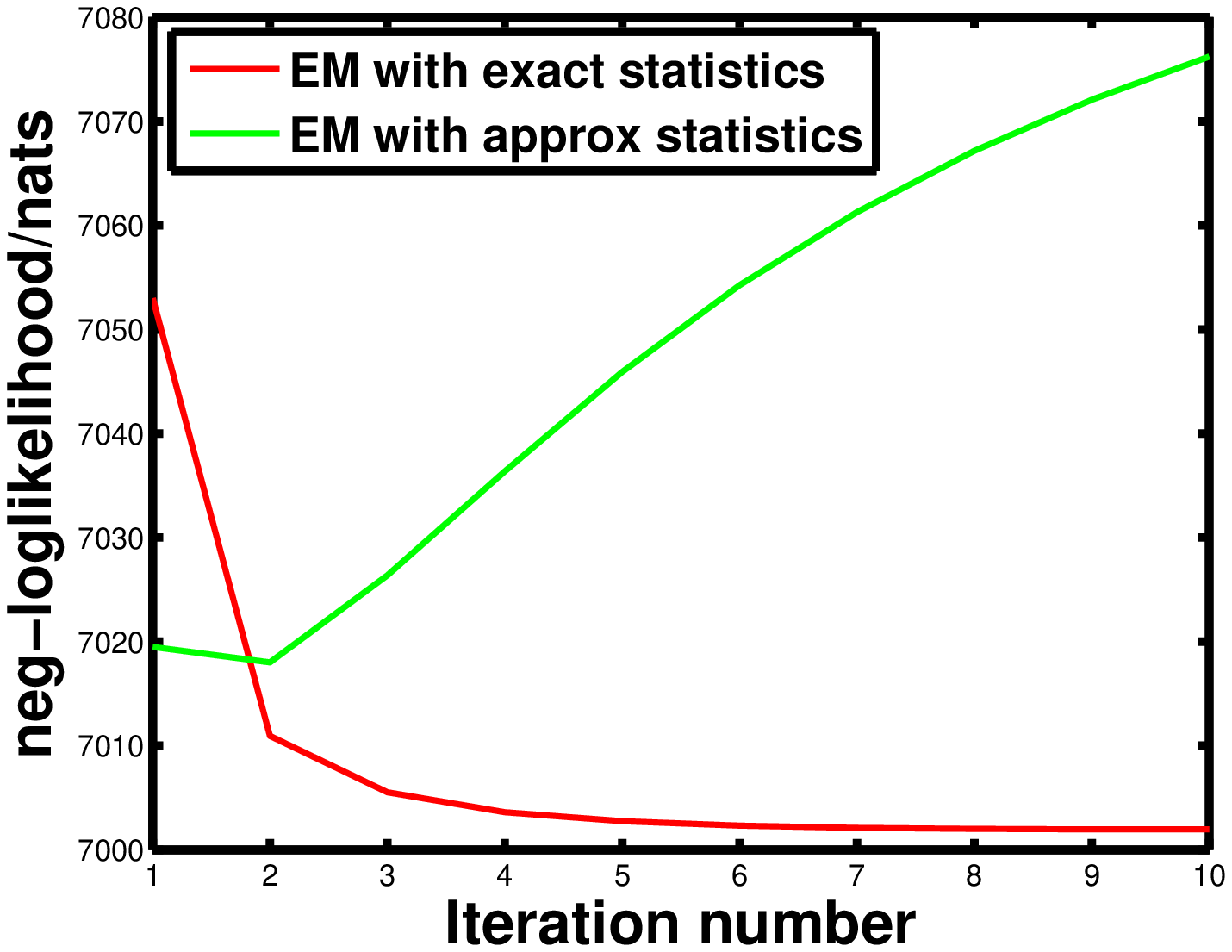}%
		\label{fig:exact-gen-jbf}}
	\vspace{-0.3cm}
	\caption{
		(a)	The negative log-likelihood of JB (EM with exact or approximated statistics) and SPLDA during training.
		(b)	The zoom-in of negative log-likelihood convergence curves for JB with exact and approximated EM statistics.
	}
	\label{fig:log-likelihood convergence of jbf and plda}
\end{figure}

As discussed before, EM iterations for SPLDA are easier to stall in a single iteration. Fig.~\ref{fig:log-likelihood convergence of jbf and plda}
 shows the neg-loglikelihood curve of SPLDA with the optimal subspace dimensionality and that of JB trained by EM with exact and approximated statistics. 
 From Fig.~\ref{fig:log-likelihood convergence of jbf and plda}, first we find that JB converges faster than SPLDA with better parameter estimation. Second, we find that JB trained by EM with approximated statistics proposed by \cite{chen2016efficient} will diverge in Fig~\ref{fig:exact-gen-jbf}, while JB trained EM with exact statistics converges nicely.

\begin{table}
	\vspace{0.2cm}
	\centering
	\scriptsize
	\begin{tabular}{l|ccc|ccc}
		\hline
		System  &\multicolumn{3}{|c|}{SRE10 MALE}&\multicolumn{3}{c}{SRE10 FEMALE} \\
		& EER & DCF10 & DCF08& EER & DCF10 & DCF08\\\hline
		LDA+COS  &1.905& 0.292&0.091 &2.619&0.399&0.126 \\
		Kaldi PLDA &1.299&0.284&0.079&1.944&0.345&0.102 \\
		SPLDA &1.010&0.217&0.055&1.621&0.287&0.079 \\
		JB&\textbf{0.894}&\textbf{0.188}&\textbf{0.048}&\textbf{1.485}&\textbf{0.245}&\textbf{0.069}\\ \hline
	\end{tabular}
	\vspace{-0.2cm}
	\caption{Performance comparison of four different discriminant analysis back-ends on NIST SRE10 core condition 5.}
	\label{table:1}
\end{table}

\begin{table}[htb]
	\vspace{0.2cm}
	\centering
	\scriptsize
	\begin{tabular}{l|c|c}
		\hline
		EER                      & SRE10 MALE     & SRE10 FEMALE   \\ \hline
		JB           &0.894           &1.485            \\
		JB-SVD (dim=403)      & 0.901          &1.513           \\
		JB-SD (dim=407)      & 0.899          &1.510          \\
		SPLDA (dim=403)   &0.981           &1.674           \\
		SPLDA (dim=407)   &0.981           &1.673           \\\hline
	\end{tabular}
	\caption{The effect of dimensionality reduction for JB and PLDA. The dimension of subspace for JB-SVD is determined by $dim=rank(A)=403$ ($A$ is defined in \cite{chen2016efficient}) and the dimension of the subspace for JB-SD is determined by $dim=rank(S_\mu)=407$. }
	\label{table:2}
	\vspace{0.1cm}
\end{table}

Tab. \ref{table:2} shows the differences on EER between SPLDA and JB with or without dimension reduction. It is observed that even with the same dimension of the subspace learned by JB, SPLDA is still worse than JB. This justifies our analysis that model formulation and hidden variable selection of JB leads to better parameter estimation than SPLDA.

\section{CONCLUSIONS}
\label{sec:typestyle}
In this paper, we propose to apply JB to model i-vectors with careful parameterization and hidden variable selection that benefits EM iterations. Both theoretical derivation and experiments conducted on the NIST SRE10 core condition demonstrate that: 1) the parameterization of JB enables it to learn the intrinsic dimensionality of the identify subspace, which can reduce the system complexity without performance degradation; 2) Hidden variables selection of JB makes EM iterations converge faster with better parameter estimation; 3) The EM with exact statistics performs better than with approximated statistics. For future work, it is interesting to apply data domain adaption \cite{garcia2014supervised} and feature compensation \cite{richardsonchannel,cumanivector} and nearest-neighbor discriminant analysis (NDA) \cite{Sadjadi2016}\cite{Fukunaga1983} that have been successfully applied to PLDA to JB to further improve performance.
\vfill\pagebreak

\bibliographystyle{IEEEbib}
\bibliography{refs,strings}

\end{document}